# Big Questions for Social Media Big Data: Representativeness, Validity and Other Methodological Pitfalls

**Zeynep Tufekci**

University of North Carolina, Chapel Hill

zeynep@unc.edu

**Abstract**

Large-scale databases of human activity in social media have captured scientific and policy attention, producing a flood of research and discussion. This paper considers methodological and conceptual challenges for this emergent field, with special attention to the validity and representativeness of social media big data analyses. Persistent issues include the over-emphasis of a single platform, Twitter, sampling biases arising from selection by hashtags, and vague and unrepresentative sampling frames. The socio-cultural complexity of user behavior aimed at algorithmic invisibility (such as subtweeting, mock-retweeting, use of "screen captures" for text, etc.) further complicate interpretation of big data social media. Other challenges include accounting for field effects, i.e. broadly consequential events that do not diffuse only through the network under study but affect the whole society. The application of network methods from other fields to the study of human social activity may not always be appropriate. The paper concludes with a call to action on practical steps to improve our analytic capacity in this promising, rapidly-growing field.

## Introduction

Very large datasets, commonly referred to as *big data*, have become common in the study of everything from genomes to galaxies, including, importantly, human behavior. Thanks to digital technologies, more and more human activities leave imprints whose collection, storage and aggregation can be readily automated. In particular, the use of social media results in the creation of datasets which may be obtained from platform providers or collected independently with relatively little effort as compared with traditional sociological methods.

*Social media big data* has been hailed as key to crucial insights into human behavior and extensively analyzed by scholars, corporations, politicians, journalists, and governments (Boyd and Crawford 2012; Lazer et al, 2009). Big data reveal fascinating insights into a variety of questions, and allow us to observe social phenomena at a previously unthinkable level, such as the mood oscillations of millions of people in 84 countries (Golder et al., 2011), or in cases where there is arguably no other feasible method of data collection, as with the study of ideological polarization on Syrian Twitter (Lynch, Freelon and Aday, 2014).

The emergence of big data from social media has had impacts in the study of human behavior similar to the introduction of the microscope or the telescope in the fields of biology and astronomy: it has produced a qualitative shift in the scale, scope and depth of possible analysis. Such a dramatic leap requires a careful and systematic examination of its methodological implications, including trade-offs, biases, strengths and weaknesses.

This paper examines methodological issues and questions of inference from social media big data. Methodological issues including the following: 1. The model organism problem, in which a few platforms are frequently used to generate datasets without adequate consideration of their structural biases. 2. Selecting on dependent variables without requisite precautions; many hashtag analyses, for example, fall in this category. 3. The denominator problem created by vague, unclear or unrepresentative sampling. 4. The prevalence of single platform studies which overlook the wider social ecology of interaction and diffusion.

There are also important questions regarding what we can legitimately infer from online imprints, which are but one aspect of human behavior. Issues include the following: 1. Online actions such as clicks, links, and retweets are complex social interactions with varying meanings, logics and implications, yet they may be aggregated together. 2. Users engage in practices that may be unintelligible to algorithms, such as subtweets (tweets referencing an unnamed but implicitly identifiable individual), quoting text via screen captures, and "hate-linking"—linking to denounce rather than endorse. 3. Network methods from other fields are often used to study human behavior without evaluating their appropriateness. 4. Social media data almost solely captures "node-to-node" interactions, while "field" effects—events that affect a society or a group in a wholesale fashion either through shared experience or through broadcast media—may often account for observed phenomena. 5. Human self-awareness needs to be taken into account; humans will alter behavior because they know they are being observed, and this change in behavior may correlate with big data metrics.

# Methodological Considerations

## 1. Model Organisms and Research: Twitter as the Field's Drosophila Melanogaster.

While there are many social media platforms, big data research focuses disproportionately on Twitter. For example, ICWSM, perhaps the leading selective conference in the field of social media, had 72 full papers last year (2013), almost half of which presented data that was primarily or solely drawn from Twitter. (Disclosure: I've long been involved in this conference in multiple capacities and think highly of it. The point isn't about any one paper's worth or quality but rather about the prevalence of attention to a single platform).

This preponderance of Twitter studies is mostly due to availability of data, tools and ease of analysis. Very large data sets, millions or billions of points, are available from this source. In contrast to Facebook, the largest social media platform, almost all Twitter activity, other than direct messages and private profiles, is visible on the public Internet. More Facebook users (estimated to be more than 50%) have made their profiles "private", i.e. not accessible on the public Internet, as compared with Twitter users (estimated as less than 10%). While Twitter has been closing some of the easier means of access, the bulk of Facebook is largely inaccessible except by Facebook's own data scientists. (Though Facebook public pages are available through its API). Unsurprisingly, only about 5% of the papers presented in ICWSM 2013 were about Facebook, and nearly all of them were co-authored with Facebook data scientists.

Twitter data also has a simple and clean structure. In contrast to the finer-grained privacy settings on other social media platforms, Twitter profiles are either "all public" or "all private." With only a few basic functions (retweet, mention, and hashtags) to map, and a maximum of 140 characters per tweet, the datasets generated by Twitter are relatively easy to structure, process and analyze as compared with most other platforms. Consequently, Twitter has emerged as a "model organism" of big data.

In biology, "model organisms" refer to species which have been selected for intensive examination by the research community in order to shed light on key biological processes such as fundamental properties of living cells. Focusing on model organisms is conducive to progress in basic questions underlying the entire field, and this approach has been spectacularly successful in molecular biology (Fields and Johnson, 2007; Geddes, 1990). However, this investigative path is not without tradeoffs.

Biases in "model-organism" research programs are not the same as sample bias in survey research, which also impact social media big data studies. The focus on just a few platforms introduces non-representativeness at the level of *mechanisms*, not just at the level of *samples*. Thus, drawing a Twitter sample that is representative of adults in the target population would not solve the problem.

To explain the issue, consider that all dominant biological model organisms – such as the fruit fly *Drosophila melanogaster*, the bacterium *Escherichia coli*, the nematode worm *Caenorhabditis elegans*, and the mouse *Mus musculus* – were selected for rapid life cycles (quick results), ease of breeding in artificial settings and small adult size (lab-friendliness), "rapid and stereotypical" development (making experimental comparisons easier), and "early separation of germ line and soma," (reducing certain kinds of variability) (Bolker, 1995; Jenner and Wills, 2007). However, the very characteristics that make them useful for studying certain biological mechanisms come at the expense of illuminating others (Gilbert, 2001, Jenner and Wills, 2007). Being easy to handle in the laboratory, in effect, implies "relative insensitivity to environmental influences" (Bolker, 1995, p:451–2) and thus unsuitability for the study of environmental interactions. The rapid development cycle depresses mechanisms present in slower-growing species, and small adult size can imply "simplification or loss of structures, the evolution of morphological novelties, and increased morphological variability" (Bolker, 1995, p:451).

In other words, model organisms can be unrepresentative of their taxa, and more importantly, may be skewed with regard to the importance of mechanisms in their taxa. They are chosen *because* they don't die easily in confinement and therefore encapsulate mechanisms specific to surviving in captivity—a trait that may not be shared with species not chosen as model organisms. The fruit fly, which breeds easily and with relative insensitivity to the environment, may lead to an emphasis on genetic factors over environmental influences. In fact, this appears to be a bias common to most model organisms used in biology.

Barbara McClintock's discovery of transposable genes, though much later awarded the Nobel prize, was initially disbelieved and disregarded partly because the organism she used, maize, was not a model organism at the time (Pray and Zhaurova, 2008). Yet the novel mechanisms which she discovered were not as visible in the more accepted model organisms, and would not have been found had she kept to those. Consequently, biologists interested in the roles of ecology and development have widened the range of organisms studied in order to uncover a wider range of mechanisms.

The dominance of Twitter as the "model organism" for social media big data analyses similarly skews analyses of *mechanisms*. Each social media platform carries with it a suite of affordances - things that it allows and makes easy versus things that are not possible or difficult - which help structure behavior on the platform. For Twitter, the key characteristics are short message length, rapid turnover, public visibility, and a directed network graph ("follow" relationships do not need to be mutual.) It lacks some of the characteristics that blogs, LiveJournal communities, or Facebook possess, such as longer texts, lengthier reaction times, stronger integration of visuals with text, the mutual nature of "friending" and the evolution of conversations over longer periods of time.

Twitter's affordances and the mechanisms it engenders interact in multiple ways. Its lightweight interface, suitable to mobile devices and accessible via texting, means that it is often the platform of choice when on the move, in low-bandwidth environments or in high-tension events such as demonstrations. The retweet mechanism also generates its own complex set of status-related behaviors and norms that do not necessarily translate to other platforms. Also, crucially, Twitter is a directed graph in that one person can "follow" another without mutuality. In contrast, Facebook's backbone is mostly an undirected graph in which "friending" is a two-way relationship and requires mutual consent. Similarly, Livejournal's core interactions tend to occur within "friends lists". Consequently, Twitter is more likely than other platforms to sustain bridge mechanisms between communities and support connections between densely interconnected clusters that are otherwise sparsely connected to each other.

To see the implications for analysis and interpretation of big data, let's look at bridging as a mechanism and consider a study which shows that bit.ly shortened links, distributed on Twitter using revolutionary hashtags during the Arab Spring, played a key role as an information conduit from within the Arab uprisings to the outside world—in other words, as bridges (Aday et al, 2012). Given its dependence on Twitter's affordances, this finding should not be generalized to mean that social media as a whole also acted as a bridge, that social media was primarily useful as a bridging mechanism, nor that Twitter was solely a bridge mechanism (since this analyzed only of tweets containing bitl.ly links). Rather, this finding speaks to a convergence of user needs and certain affordances in a subset of cases which fueled one mechanism: in this case, bridging.

Finally, there is indeed a sample bias problem. Twitter is used by less than 20% of the US population (Mitchell and Hitlin, 2014), and that is not a randomly selected group. While Facebook has wider diffusion, its use is also structured by race, gender, class and other factors (Hargittai, 2008). Thus, the use of a few online platforms as "big data" model organisms raises important questions of representation and visibility, as different demographic or social groups may have different behavior—online and offline—and may not be fully represented or even sampled via current methods.

All this is not to say that Twitter is an inappropriate platform to study. Research in the model organism paradigm allows a large community to coalesce around shared datasets, tools and problems, and can produce illuminating results. However, the specifics of the "model organism" and biases that may result should not be overlooked.

## 2. Hashtag Analyses, Selecting on the Dependent Variable, Selection Effects and User Choices.

The inclusion of hashtags in tweets is a Twitter convention for marking a tweet as part of a particular conversation or topic, and many social media studies rely on them for sample extraction. For example, the Tunisian uprising was associated with the hashtag #sidibouzid while the initial Egyptian protests of January 25, 2011, with #jan25. Facebook's adoption of hashtags makes the methodological specifics of this convention even more important. While hashtag studies can be a powerful for examining network structure & information flows, all hashtag analyses, by definition, select on a dependent variable, and hence display the concomitant features and weaknesses of this methodological path.

"Selecting on the dependent variable" occurs when inclusion of a case in a sample depends on the very variable being examined. Such samples have specific limits to their analytic power. For example, analyses that only examine revolutions or wars that have *occurred* will overlook cases where the causes and correlates of revolution and war have been present but in which there have been no resulting wars or revolutions (Geddes, 2010). Thus, selecting on the dependent variable (the occurrence of war or revolution) can help identify necessary conditions, but those may not be sufficient. Selecting on the dependent variable can introduce a range of errors specifics of which depend on the characteristics of the uncorrelated sample.

In hashtag datasets, a tweet is included because the user chose to use it, a clear act of self-selection. Self-selected samples often will not only have different overall characteristics than the general population, they may also exhibit significantly different correlational tendencies which create thorny issues of confounding variables. Famous examples include the hormone replacement therapy (HRT) controversy in which researchers had, erroneously, believed that HRT conferred health benefits to post-menopausal women based on observational studies of women who self-selected to take HRT. In reality, HRT therapy was adopted by healthier women. Later randomized double-blind studies showed that HRT was, in fact, harmful—so harmful that the researchers stopped the study in its tracks to reverse advice that had been given to women for a decade.

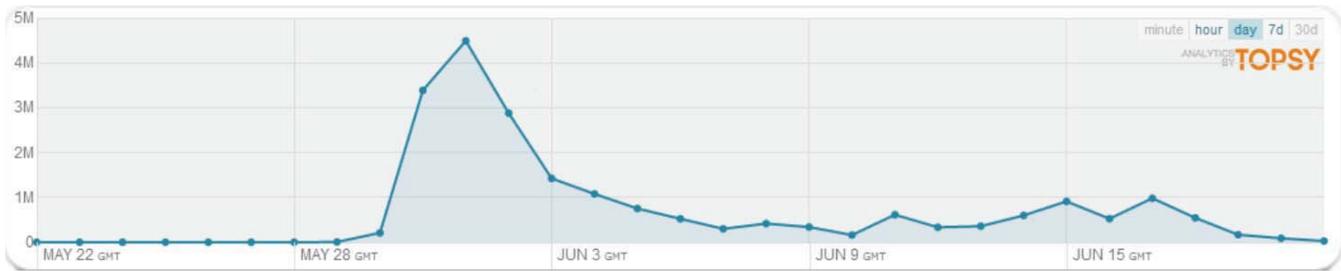

*Figure 1: The frequency of top 20 hashtags associated with Gezi Protests. (Banko and Babacan, 2013)*

Samples drawn using different hashtags can differ in important dimensions, as hashtags are embedded in particular cultural and socio-political frameworks. In some cases, the hashtag is a declaration of particular sympathy. In other cases, there may be warring messages as the hashtag emerges as a contested cultural space. For example, two years of regular monitoring of activity—checking at least for an hour once a week—on the hashtags #jan25 and #Bahrain show their divergent nature. Those who choose to use #jan25 are almost certain to be sympathetic to the Egyptian revolution while #Bahrain tends to be used both by supporters and opponents of the uprising in Bahrain. Data I systematically sampled on three occasions showed that only about 1 in 100 #jan25 tweets were neutral while the rest were all supporting the revolution. Only about 5 out of 100 #Bahrain tweets were neutral, and 15 out of 100 were strongly opposed to the uprising, while the rest, 80 out of 100 were supportive. In contrast, #cairotraffic did not exhibit any overt signs of political preference. Consequently, since the hashtag users are a particular community, thus prone to selection biases, it would be difficult to generalize from their behavior to other samples. Political users may be more prone to retweeting, say, graphic content, whereas non-political users may react with aversion. Hence, questions such as "does graphic content spread quickly on Twitter" or "do angry messages diffuse more quickly" might have quite different answers if the sample is drawn through different hashtags.

Hashtag analyses can also be affected by user activity patterns. An analysis of twenty hashtags used during the height of Turkey's Gezi Park protests in June 2013 (#occupygezi, #occupygeziparki, #direngeziparki, #direnankara, #direngaziparki, etc.) shows a steep rise in activity on May 30th when the protests began, dropping off by June 3rd (Figure 1). Looking at this graph, one might conclude that either the protests had died down, or that people had stopped talking about the protests on Twitter. Both conclusions would be very mistaken, as revealed by the author's interviews with hundreds of protesters on-the-ground during the protests, online ethnography that followed hundreds of people active in the protests (some of them also interviewed offline), monitoring of Twitter, trending topics, news coverage and the protests themselves.

What had happened was that as soon as the protest became the dominant story, large numbers of people continued to discuss them heavily – almost to the point that no other discussion took place on their Twitter feeds – but stopped using the hashtags except to draw attention to a new phenomenon or to engage in "trending topic wars" with ideologically-opposing groups. While the protests continued, and even intensified, the hashtags died down. Interviews revealed two reasons for this. First, once everyone knew the topic, the hashtag was at once superfluous and wasteful on the character-limited Twitter platform. Second, hashtags were seen only as useful for attracting attention to a particular topic, not for talking about it.

In August, 2013, a set of stairs near Gezi Park which had been painted in rainbow colors were painted over in drab gray by the local municipality. This sparked outrage as a symbolic moment, and many people took to Twitter under the hashtag #direnmerdiven (roughly "#occupystairs"). The hashtag quickly and briefly trended and then disappeared from the trending list as well as users' Twitter streams. However, this would be a misleading measure of activity on the painting of the stairs, as monitoring a group who had been using the hashtag showed that almost all of them continued to talk about the topic intensively on Twitter, but without the hashtag. Over the next week, hundreds, maybe thousands of stairs in Turkey were painted in rainbow colors as a form of protest, a phenomenon not at all visible in any data drawn from the hashtag.

Finally, most hashtags used to build big datasets are successful hashtags - ones that got well-known, distributed widely and generated large amount of interest. It is likely that the dynamics of such events differ significantly from those of less successful ones. In sum, hashtag datasets should be seen as self-selected samples with data "missing not at random" and interpreted accordingly (Allison, 2001; Meiman and Freund, 2012; Outhwaite et al, 2007)

All this is not to argue that hashtag datasets are not useful. In contrast, they can provide illuminating glimpses into specific cultural and socio-political conversations. However, hashtag dataset analyses need to be accompanied by a thorough discussion of the culture surrounding the specific hashtag, and analyzed with careful consideration of selection and sampling biases.

There might be ways to structure the sampling of Twitter datasets so that the hashtag is not the sole criterion. For example, Freelon, Lynch and Aday (2014) extracted a dataset first based on the use of the word "Syria" in Arabic or English, and then extracted hashtags from that dataset while also performing analyses on the wider dataset. Another method might be to use the hashtag to identify a sample of users and then collect tweets of those users (who will likely drop using the hashtag) rather than collecting the tweets via the hashtag.

Above all, hashtag analyses should start from the principle of understanding user behavior first, and should follow the user rather than following the hashtag.

### 3. The Missing Denominator: We Know Who Clicked But We Don't Know Who Saw Or Could:

One of the biggest methodological dangers of big data analyses is insufficient understanding of the denominator. It's not enough to know how many people have "liked" a Facebook status update, clicked on a link, or "retweeted" a message without knowing how many people saw the item and chose not to take any action. We rarely know the characteristics of the sub-population that sees the content even though that is the group, and not the entire population, from which we are sampling. Normalization is rarely done, or may even be actively decided against because the results start appearing more complex or more trivial (Cha, 2008).

While the denominator is often not calculable, it may be possible to estimate. One measure might be "potential exposure," corresponding to the maximum number of people who may have seen a message. However, this highlights another key issue: the data is often proprietary (Boyd and Crawford, 2012). It might be possible to work with the platforms to get estimates of visibility, click-through and availability. For example, Facebook researchers have disclosed that the mean and median fraction of a user's friends that see status update posts is about 34 to 35%, though the distribution of the variable seems to have a large spread (Bernstein et al., 2013).

With some disclosure from proprietary platforms, it may be possible to calculate "likely" exposure numbers based on "potential" exposure - similar to the way election polls model "likely" voters or TV ratings try to measure people watching a show rather than just being in the room where the TV is on. Steps in this direction are likely to be complex and difficult, but without such efforts, our ability to interpret raw numbers will remain limited. The academic community should ask for more disclosure and access from the commercial platforms.

It's also important to normalize underlying populations when comparing "clicks," "links," or tweets. For example, Aday et al. (2012) compares numbers of clicks on bit.ly links in tweets containing hashtags associated with the Arab uprisings and concludes that "new media outlets that that use bit.ly are more likely to spread information outside the region than inside it." This is an important finding. However, interpretation of this finding should take into account the respective populations of Twitter users in the countries in question. Egypt's population is about 80 million, about 1 percent of the global population. Any topic of global interest about Egypt could very easily generate more *absolute* number of clicks outside the country even if the activity within the country remained much more concentrated in *relative* proportions. Finally, the size of these datasets makes traditional measures like statistical significance less valuable (Meiman and Freund, 2012), a problem exacerbated by lack of information about the denominator.

### 4. Missing the Ecology for the Platform:

Most existing big data analyses of social media are confined to a single platform (often Twitter, as discussed.) However, most of the topics of interest in such studies, such as influence or information flow, can rarely be confined to the Internet, let alone to a single platform. The difficulty in obtaining high-quality multi-platform data does not mean that we can treat a single platform as a closed and insular system. Information in human affairs flows through all available channels.

The emergent media ecology is a mix of old and new media which is not strictly segregated by platform or even by device. Many "viral" videos take off on social media only after being featured on broadcast media, which often follows their being highlighted on intermediary sites such as Reddit or Buzzfeed. Political news flowing out of Arab Spring uprisings to broadcast media was often curated by sites such as Nawaat.org that had emerged as trusted local information brokers. Analysis from Syria shows a similar pattern (Aday et al. 2014). As these examples show, the object of analysis should be this integrated ecology, and there will be significant shortcomings in analyses which consider only a single platform.

Link analyses on hashtags datasets for the Arab uprisings show that the most common links from social media are to the websites of broadcast media (Aday et al. 2012). The most common pattern was that users alternate between Facebook, Twitter, broadcast media, cell-phone conversations, texting, face-to-face and other methods of interaction and information sharing (Tufekci & Wilson, 2012).

These challenges do not mean single-platform analyses are not valuable. However, all such analyses must take into account that they are not examining a closed system and that there may be effects which are not visible because the relevant information is not contained within that platform. Methodologically, single-platform studies can be akin to looking for our keys under the light. More research, admittedly much more difficult and expensive than scraping data from one platform, is needed to understand broader patterns of connectivity. Sometimes, the only way to study people is to study people.

## Inferences and Interpretations

The question of inference from analyses of social media big data remains underconceptualized and underexamined. What's a click? What does a retweet mean? In what context? By whom? How do different communities interpret these interactions? As with all human activities, interpreting online imprints engages layers of complexity.

### 1. What's in a Retweet? Understanding our Data:

The same act can have multiple, even contradictory meanings. In many studies, for example, retweets or mentions are used as proxies for influence or agreement. This may hold in some contexts; however, there are many conceptual steps and implicit assumptions embedded in this analysis. It is clear that a retweet is information exposure and/or reaction; however, after that, its meaning could range from affirmation to denunciation to sarcasm to approval to disgust. In fact, many social media acts which are designed as "positive" interactions by the platform engineers, ranging from Retweets on Twitter to even "Likes" on Facebook can carry a range of meanings, some quite negative.

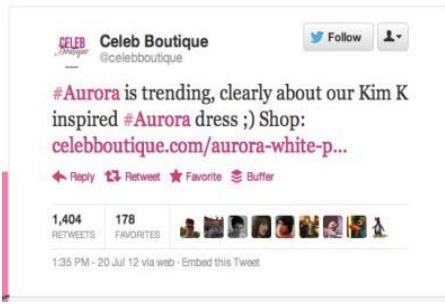

*Figure 2: Retweeted widely, but mostly in disgust*

As an example, take the recent case of the twitter account of fashion store @celebboutique. On July, 2012, the account tweeted with glee that the word "#aurora" was trending and attributed this to the popularity of a dress named #aurora in its shop. The hashtag was trending, however, because Aurora, Colorado was the site of a movie theatre massacre on that day. There was an expansive backlash against @celebboutique's crass and insensitive tweet. There were more than 200 mentions and many hundreds of retweets with angry messages in as little as sixty seconds. The tweet itself, too, was retweeted thousands of times (See Figure 2). After about an hour, the company realized its mistake and stepped in. This was followed by more condemnation—a few hundred mentions per minute at a minimum. (For more analysis: (Gilad, 2012)) Hence, without understanding the context, the spike in @celebboutique mentions could easily be misunderstood.

Polarized situations provide other examples of "negative retweets." For example, during the Gezi protests in Turkey, the mayor of Ankara tweeted personally from his account, often until late hours of the night, engaging Gezi protesters individually in his idiosyncratic style, which involved the use of "ALL CAPS" and colorful language. He became highly visible among supporters as well as opponents of these protests. His visibility, combined with his style, meant that his tweets were widely retweeted—but not always by supporters. Gezi protesters would retweet his messages and then follow the retweet with a negative or mocking message. His messages were also retweeted without comment by people whose own Twitter timelines made clear that their intent was to "expose" or ridicule, rather than agree. A simple aggregation would find that thousands of people were retweeting his tweets, which might be interpreted as influence or agreement.

One of the most cited Twitter studies (Kwak et al.) grapples with how to measure influence, and asks whether the number of followers or the number of retweets is a better measure. That paper settles on retweets, stating that "The number of retweets for a certain tweet is a measure of the tweet's popularity and in turn of the tweet writer's popularity." The paper then proceeds to rank users by the total number of retweets, and refers to this ranking alternatively as influence or popularity. Another important social media study, based on Twitter, speaks of in-degree (number of followers) as a user's popularity, and retweets as influence (Cha et al., 2010). Both are excellent studies of retweet and following behavior, but in light of the factors discussed above, "influence" and "popularity" are may not be the best term to use for the variables they are measuring. Some portion of retweets and follows are, in fact, negative or mocking, and do not represent "influence" in the way it is ordinarily understood. The scale of such behavior remains an important, unanswered question (Freelon, 2014).

### 2. Engagement Invisible to Machines: Subtweets, Hate-Links, Screen Captures and Other Methods:

Social media users engage in practices that alter their visibility to machine algorithms, including *subtweeting*, discussing a person's tweets via "*screen captures,*" and *hate-linking*. All these practices can blind big data analyses to this mode of activity and engagement.

*Subtweeting* is the practice of making a tweet referring to a person algorithmically invisible to that person—and consequently to automated data collection—even as the reference remains clear to those "in the know." This manipulation of visibility can be achieved by referring to a person who has a twitter handle without either "mentioning" this handle, or by inserting a space between the @

sign and the handle, or by using their regular name or a nickname rather than the handle, or even sometimes deliberately misspelling the name. In some cases, the reference can only be understood in context, as there is no mention of the target in any form. These many forms of subtweeting come with different implications for big data analytics.

For example, a controversial article by Egyptian-American Mona El Eltahawy sparked a massive discussion in Egypt's social media. In a valuable analysis of this discussion, sociologists Alex Hanna and Marc Smith extracted the tweets which mentioned her or linked to the article. Their network analysis revealed great polarization in the discussion, with two distinctly clustered groups. However, while watching this discussion online, I noticed that many of the high-profile bloggers and young Egyptian activists discussing the article - and greatly influencing the conversation - were indeed subtweeting. Later discussions with this community revealed that this was a deliberate choice that had been made because many people did not want to "give Eltahawy attention," even as they wanted to discuss the topic and her work.

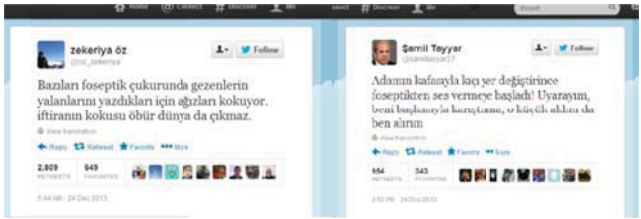

*Figure 3: Two people "subtweeting" each other without mentioning names. The exchange was clear enough, however, to be reported in newspapers.*

In another example drawn from my primary research on Turkey, figure 3 shows a subtweet exchange between two prominent individuals that would be unintelligible to anyone who did not already follow the broader conversation and was not intimately familiar with the context. While each person is referring to the other, there are no names, nicknames, or handles. In addition, neither follows the other on Twitter. It is, however, clearly a direct engagement and conversation, if a negative one. A broad discussion of this "Twitter spat" on Turkish Twitter proved people were aware of this as a two-way conversation. It was so well understood that it was even reported in newspapers.

While the true prevalence of this behavior is hard to establish, exactly because the activity is hidden from large-scale, machine-led analyses, observations of Turkish Twitter during the Gezi protests of June 2013 revealed that such subtweets were common. In order to get a sense of its scale, I undertook an online ethnography in December, 2013, during which two hundred Twitter users from Turkey, assembled as a purposive sample including ordinary users as well as journalists and pundits, were followed for an hour at a time in, totaling at least 10 hours of observation dedicated to catching subtweets. This resulted in a collection of 100 unmistakable subtweets; many more were undoubtedly missed because they are not always obvious to observers. In fact, the subtweets were widely understood and retweeted, which increases the importance of such practices. Overall, the practice appears common enough to be described as routine, at least in Turkey.

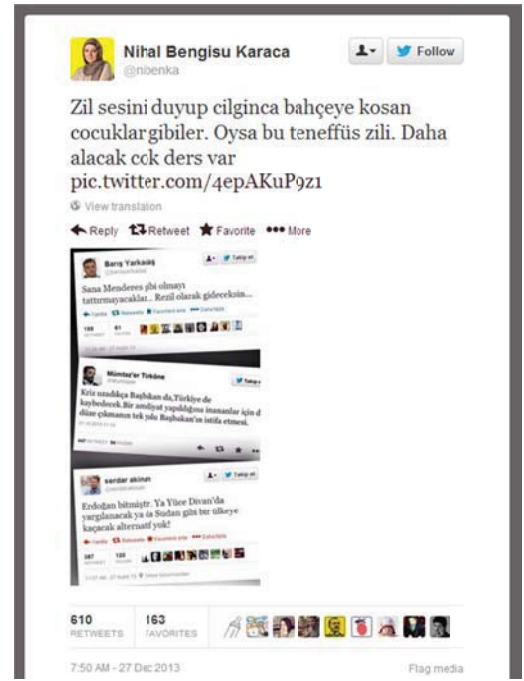

*Figure 4: Algorithmically Invisible Engagement: A columnist responds to critics by screen captures.*

Using screen captures rather than quotes is another practice that adds to the invisibility of engagement to algorithms. A "caps" is done when Twitter users reference each other's tweets through screen captures rather than links, mentions or quotes. An example is shown on Figure 4. This practice is so widespread that a single hour following the same purposive sample resulted in more than 300 instances in which users employed such "caps."

Yet another practice, colloquially known as "hate-linking," limits the algorithmic visibility of engagement, although this one is potentially traceable. "Hate-linking" occurs when a user links to another user's tweet rather than mentioning or quoting the user. This practice, too, would skew analyses based on mentions or retweets, though in this case, it is at least possible to look for such links.

Subtweeters, "caps" users, and hate-linkers are obviously a smaller community than tweeters as a whole. While it is unclear how widespread these practices truly are, studying Turkish Twitter shows that they are not uncommon, at least in that context. Other countries might have specific social media practices that confound big data analytics in different ways. Overall, a simple "scraping" of Turkish Twitter might produce a polarized map of groups not talking to each other, whereas the reality is a polarized situation in which contentious groups are engaging each other but without the conventional means that make such conversations visible to algorithms and to researchers.

### 3. Limits of Methodological Analogies and Importing Network Methods from Other Fields:

Do social media networks operate through similar mechanisms as networks of airlines? Does information work the way germs do? Such questions are rarely explicitly addressed even though many papers import methods from other fields on the implicit assumption that the answer is a yes. Studies that do look at this question, such as Romero et al. (2011) and Lerman et al. (2010), are often limited to single, or few, platforms, which limit their explanatory power because information among humans does not diffuse in a single platform whereas viruses do, indeed, diffuse in a well-defined manner. To step back further, even representing social media interactions as a network requires a whole host of implicit and important assumptions that should be considered explicitly rather than assumed away (Butts, 2009).

Epidemiological or contagion-inspired analyses often treat connected edges in social media networks as if they were "neighbors" in physical proximity. In epidemiology, it is reasonable to treat "physical proximity" as a key variable, assuming that adjacent nodes are "susceptible" to disease transmission for very good reason: the underlying model is a well-developed, empirically-verified germ-theory of disease in which small microbes travel in actual space (where distance matters) to infect the next person by entering their body. This physical process has well-understood properties, and underlying probabilities can often be calculated with precision.

Creating an analogy from social media interactions to physical proximity may be a reasonable and justified under certain conditions and assumptions, but this step is rarely subjected to critical examination (Salathé et al, 2013). There are significant differences between germs and information traveling in social media networks. Adjacency in social media is multi-faceted; it cannot always be mapped to physical proximity; and human "nodes" are subject to information from a wide range of sources, not just those they are connected to in a particular social media platform. Finally, whether there is a straightforward relation between information exposure and the rate of "influence," as there often is for exposure to a disease agent and the rate of infection, is something that should be empirically investigated, not assumed.

There are clearly similar dynamics in different types of networks, human and otherwise, and the different fields can learn much from each other. However, importation of methods needs to rely on more than some putative universal, context-independent property of networked interaction simply by virtue of the existence of a network.

### 4. Field Effects: Non-Networks Interactions

Another difference between spatial or epidemiological networks and human social networks is that human social information flows do not occur only through node-to-node networks but also through field effects, large-scale societal events which impact a large group of actors contemporaneously. Big events, weather occurrences, etc. all have society-wide field effects and often do not diffuse solely through interpersonal interaction (although they also do greatly impact interpersonal interaction by affecting the agenda, mood and disposition of individuals).

For example, most studies agree that Egyptian social media networks played a role in the uprising which began in Egypt in January 2011 and were a key conduit of protest information (Lynch, 2012; Aday et al, 2012; Tufekci and Wilson, 2012). However, there was almost certainly another important information diffusion dynamic. The influence of the Tunisian revolution itself on the expectations of the Egyptian populace was a major turning point (Ghonim, 2012; Lynch, 2012). While analysis of networks in Egypt might not have revealed a major difference between the second and third week of January of 2011, something major had changed in the *field*. To translate it into epidemiological language, due to the Tunisian revolution and the example it presented, *all* the "nodes" in the network had a different "susceptibility" and "receptivity" to information about an uprising. The downfall of the Tunisian president, which showed that even an enduring autocracy in the Middle East was susceptible to street protests, energized the opposition and changed the political calculation in Egypt. This information was diffused through multiple methods and broadcast media played a key role. Thus, the communication of the Tunisia effect to the Egyptian network was not necessarily dependent on the network structure of social media.

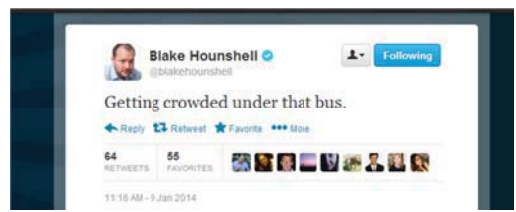

*Figure 5: Clear meaning only in context and time.*

Social media itself is often incomprehensible without reference to field events outside it. For example, take the tweet in Figure 5. The tweets merely states: "Getting crowded under that bus." Strangely, it has been tweeted more than sixty times and favorited more than 50. For those following in real time, this was an obvious reference to New Jersey Governor Chris Christie's press conference in which he blamed multiple aides for the closing of a bridge which caused massive traffic jams, allegedly to punish a mayor who did not endorse him. Without understanding the Chris Christie press conference, neither the tweet, nor many retweets of it are interpretable.

The turn to networks as a key metaphor in social sciences, while fruitful, should not diminish our attention to the multi-scale nature of human social interaction

### 5. You Name It, Humans Will Game it: Reflexivity and Humans:

Unlike disease vectors or gases in a chamber, humans understand, evaluate and respond to the same metrics that big data researchers are measuring. For example, political activists, especially in countries such as Bahrain, where the unrest and repression have received less mainstream global media attention, often undertake deliberate attempts to make a hashtag "trend." Indeed, "to trend" is increasingly used as a transitive verb among these users, as in: "let's trend #Hungry4BH". These efforts are not mere blind stabs at massive tweeting; they often display a fine-tuned understanding of Twitter's trending topics algorithm, which, while not public, is widely understood through reverse engineering (Lotan, 2012). In coordinated campaigns, Bahrain activists have trended hashtags such as #100strikedays, #Bloodyf1, #KillingKhawaja, #StopTearGasBahrain, and #F1DontRaceBahrain, among others.

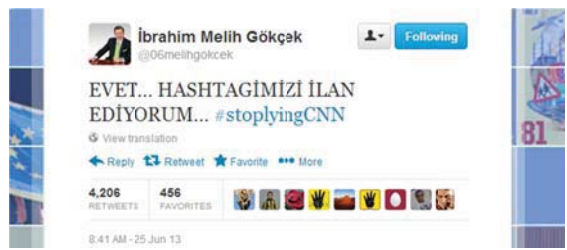

*Figure 6: Ankara Mayor leads a hashtag campaign that will eventually trend worldwide. [Translation: Yes… I'm announcing ur hashtag. #stoplyingCNN]*

Campaigns to trend hashtags are not limited to grassroots activists. In Figure 6, drawn from my primary research in Turkey, you can see AKP's Ankara mayor, an active figure in Turkish Twitter discussed before, announcing the hashtag that will be "trended": #cnnislying. This was retweeted more than 4000 times. He had been announcing the campaign and had asked people to be in front of their devices at a set time; in these campaigns, the actual hashtag is often withheld until a pre-agreed time so as to produce a maximum spike which Twitter's algorithm is sensitive to. That hashtag indeed trended worldwide. Similar coordinated campaigns are common in Turkey and occurred almost every day during the contentious protests of June, 2013.

Such behaviors, aimed at avoiding detection, amplifying a signal, or other goals, by deliberate gaming of algorithms and metrics, should be expected in all analyses of human social media. Currently, many studies do take into account "gaming" behaviors such as spam and bots; however, coordinated or active attempts by actual people to alter metrics or results, which often can only be discovered through qualitative research, are rarely taken into account.

## Conclusion: Practical Steps—a Call to Action

Social media big data is a powerful addition to the scientific toolkit. However, this emergent field needs to be placed on firmer methodological and conceptual footing. Meaning of social media imprints, context of human communications, and nature of socio-cultural interactions are multi-faceted and complex. People's behavior differs in significant dimensions from other objects of network analyses such as germs or airline flights.

The challenges outlined in this paper are results of systematic qualitative and quantitative inquiry over two years; however, this paper cannot be conclusive in identifying the scale and the scope these issues—in fact, the challenge is exactly that such issues are resistant to large-scale machine-analyses that allows us to exactly pinpoint them. While the challenges are thorny, there are many practical steps. These include:

**1- Target non-social dependent variables.** No data source is perfect, and every dataset is imperfect in its own ways. Most big data analyses remains within the confines the dataset, with little means to probe validity. Seeking dependent variables outside big data, especially those for which there are other measures obtained using traditional, tested and fairly reliable methods, and looking at the convergent and divergent points, will provide much needed clarity to strengths and weaknesses of these datasets. Such dependent variables could range from , elections results to unemployment numbers.

**2- Qualitative pull-outs.** Researchers can include a "qualitative pull-out" from their sample to examine variations in behavior. For example, what percent of retweets are "hate-retweets"? A small random subsample can provide a check. This may involve asking questions to people. For example: did the some people hear of X from TV as well as from Twitter? These qualitative pull-outs need not be huge to help interpretation.

**2. Baseline panels.** Establishing a panel to study peoples' digital behavior, similar to panel studies in social sciences, could develop "baselines" and "guidelines" for the whole community. Data sought could include those in this paper.

**3. Industry Outreach.** The field should solicit cooperation from the industry for data such as "denominators", similar to Facebook's recent release of what percent of a Facebook network sees status updates. Industry scientists who participate in the research community can be conduits.

**4. Convergent answers and complimentary methods.** Multi-method, multi-platform analyses should be sought and rewarded. As things stand, these exist (Adar et al., 2007 or Kairam, 2013) but are rare. Whenever possible, social media big data studies should be paired with surveys, interviews, ethnographies, and other methods so that biases and short-comings of each method can be used to balance each other to arrive at richer answers.

**5. Multi-disciplinary teams.** Scholars from fields where network methods are shared should cooperate to study the scope, differences and utility of common methods.

**6. Methodological awareness in review.** These issues should be incorporated into the review process and go beyond soliciting "limitations" sections.

A future study that recruited a panel of ordinary users, from multiple countries, and examined their behavior online and offline, and across multiple platforms to detect the frequency of behaviors outlined here, and those not detected yet, would be a path-breaking next step for understanding and grounding our social media big data.